B. Tech Project Progress Report

On

# Development of Adaptive Frame Reservation Scheme and Naïve Persistent State Co-Located Coexistence Controller

As

"*Solutions to WiMAX-WiFi Coexistence*"

Submitted in partial fulfillment of requirements
for the degree of

*Bachelor of Technology*
By

**Jatin Sharma [Y07UC048]**
Dept. of Communication & Computer Engineering,
The LNMIIT, Jaipur

Under the able guidance of
**Prof. V. Sinha**

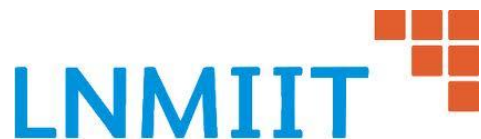

**The LNM Institute of Information Technology**
**Deemed University**
**Jaipur (Rajasthan)**

# Certificate

This is to certify that the work contained in this B.Tech Project Progress Report, entitled "Development of Adaptive Frame Reservation Scheme and Naïve Persistent State Co-Located Coexistence Controller", has been carried out by **Jatin Sharma** (Y07UC048) under my supervision during the period of August 1, 2010- December 26, 2010 and is still in progress. The algorithms, emulations and concepts implemented in this report have not been submitted to any other Institute or University for the award of any degree or diploma.

<div style="text-align: right;">

**V. Sinha**
Distinguished Professor
**LNM IIT Jaipur**
(Deemed University)

</div>

# Abstract

Future broadband networks will need to provide very high capacity at low network cost with increased revenue through enhanced or new services. Huge research has been made and WiMAX came up as one of the leading technologies which is still in its experimental phase and expected to come and capture the market soon. However the 2.3 GHz and 2.5 GHz frequency bands allocated to WiMAX falls adjacent to the unlicensed 2.4 GHz ISM band and thus creates two serious coexistence issues.

First problem is to address radio interfaces that are located on two independent platforms and still possess the potential for mutual interference owing to close proximity to each other. Such a scenario requires OVER-THE-AIR (OTA) coordination which is difficult to handle. The *'Adaptive Frame Reservation Scheme'* presented here extends the CTS frame reservation signaling defined in 802.11 specifications to a demand based and adaptive scheme. The scheme assumes future WiMAX nodes to be equipped with co-located WiFi interface as well because of its popularity and cost effectiveness. A CTS packet is sent by this collocated WiFi interface to reserve a slot. Dynamic Medium Acquisition [DMA], Adaptive CTS Power [ACP] and Dynamic Performance Evaluation [DPE] phases improves this both in terms of fairness and throughput.

Second issue is to address the coexistence problem in multi-radio platforms where two or more radios are co-located, creating an even worse interference scenario. This can be managed by hardware signaling that can be made available between radio interfaces through OS control. The development of a smart Co-located Coexistence Controller is explored which continuously receives transmission, reception and sleep requests from attached interfaces and in return grant permissions. A primitive *'Naïve Persistent State Controller'* is presented here, which completely eliminates the possibility of co-located interference. Few refinements over this Naïve controller are suggested and are the scope for future work.

# Table of Contents

# Chapter 1
# WiMAX-WiFi Coexistence

**1 Introduction**

WiMAX refers to interoperable implementations of the IEEE 802.16 wireless-networks standard (ratified by the WiMAX Forum), in similarity with Wi-Fi, which refers to interoperable implementations of the IEEE 802.11 WirelessLAN standard (ratified by the Wi-Fi Alliance). The WiMAX Forum certification allows vendors to sell their equipment as WiMAX (Fixed or Mobile) certified, thus ensuring a level of interoperability with other certified products, as long as they fit the same profile.

The IEEE 802.16 standard forms the basis of 'WiMAX' and is sometimes referred to colloquially as "WiMAX", "Fixed WiMAX", "Mobile WiMAX", "802.16d" and "802.16e.".Clarification of the formal names is as follow:

- 802.16-2004 is also known as 802.16d, which refers to the working party that has developed that standard. It is sometimes referred to as "Fixed WiMAX," since it has no support for mobility.
- 802.16e-2005, often abbreviated to 802.16e, is an amendment to 802.16-2004. It introduced support for mobility, among other things and is therefore also known as "Mobile WiMAX".

Mobile WiMAX is the WiMAX incarnation that has the most commercial interest to date and is being actively deployed in many countries. Mobile WiMAX is also the basis of future revisions of WiMAX.

**1.1 Physical Layer**

The original version of WiMAX uses scalable orthogonal frequency-division multiple access (SOFDMA) as opposed to the fixed orthogonal frequency-division multiplexing (OFDM) version with 256 sub-carriers (of which 200 are used) in 802.16d. More advanced versions, including 802.16e, also bring multiple antenna support through MIMO (See WiMAX MIMO). This brings potential benefits in terms of coverage, self installation, power consumption, frequency re-use and bandwidth efficiency.

## 1.2 MAC (Data Link) Layer

The WiMAX MAC uses a scheduling algorithm for which the subscriber station needs to compete only once for initial entry into the network. After network entry is allowed, the subscriber station is allocated an access slot by the base station. The time slot can enlarge and contract, but remains assigned to the subscriber station, which means that other subscribers cannot use it. In addition to being stable under overload and over-subscription, the scheduling algorithm can also be more bandwidth efficient. The scheduling algorithm also allows the base station to control Quality of service (QoS) parameters by balancing the time-slot assignments among the application needs of the subscriber stations.

## 2 Comparison with Wi-Fi

Comparisons and confusion between WiMAX and Wi-Fi are frequent because both are related to wireless connectivity and Internet access.

1. WiMAX is a long range system, covering many kilometers that uses licensed or unlicensed spectrum to deliver connection to a network, in most cases the Internet.
2. Wi-Fi uses unlicensed spectrum to provide access to a local network.
3. Wi-Fi is more popular in end user devices.
4. Wi-Fi runs on the Media Access Control's CSMA/CA protocol, which is connectionless and contention based, whereas WiMAX runs a connection-oriented MAC.
5. WiMAX and Wi-Fi have quite different Quality of Service (QoS) mechanisms:
6. WiMAX uses a QoS mechanism based on connections between the base station and the user device. Each connection is based on specific scheduling algorithms.
7. Wi-Fi uses contention access - all subscriber stations that wish to pass data through a wireless access point (AP) are competing for the AP's attention on a random interrupt basis. This can cause subscriber stations distant from the AP to be repeatedly interrupted by closer stations, greatly reducing their throughput.
8. Both 802.11 and 802.16 define Peer-to-Peer (P2P) and ad hoc networks, where an end user communicates to users or servers on another Local Area Network (LAN) using its access point or base station. However, 802.11 supports also direct ad hoc or peer to peer networking between end user devices without an access point while 802.16 end user devices must be in range of the base station.

Wi-Fi and WiMAX are complementary. WiMAX network operators typically provide a WiMAX Subscriber Unit which connects to the metropolitan WiMAX network and provides Wi-Fi within the home or business for local devices (e.g. Laptops, Wi-Fi Handsets, smartphones) for connectivity. This enables the user to place the WiMAX Subscriber Unit in the best reception area (such as a window), and still be able to use the WiMAX network from any place within their residence.

## 3 WiMAX-WiFi Coexistence Problem

WiMAX [Worldwide Interoperability for Microwave Access] is the most promising high data rate standard enabling the delivery of last mile wireless broadband access as an alternative to cable and DSL. WiMAX works at 2.3 GHz and 2.5 GHz, which are adjacent to license exempt 2.4 GHz ISM band.

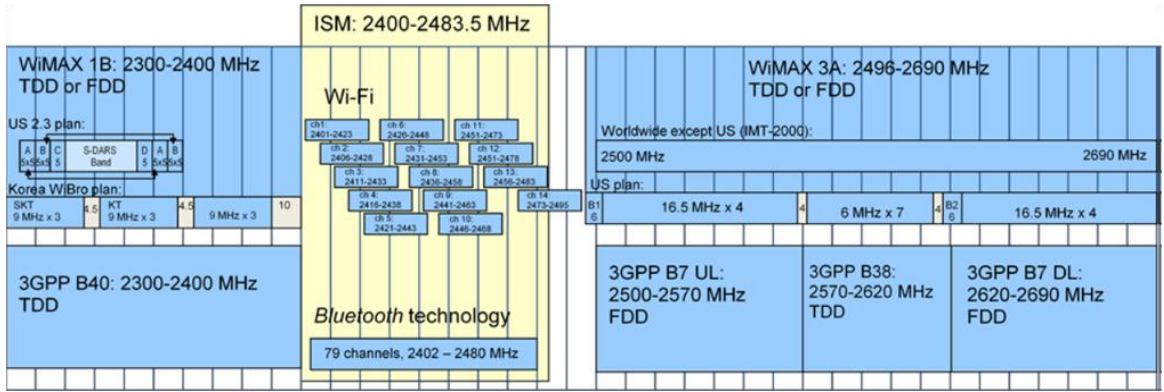

Figure 1.1: Frequency allocation near 2.4 GHz.

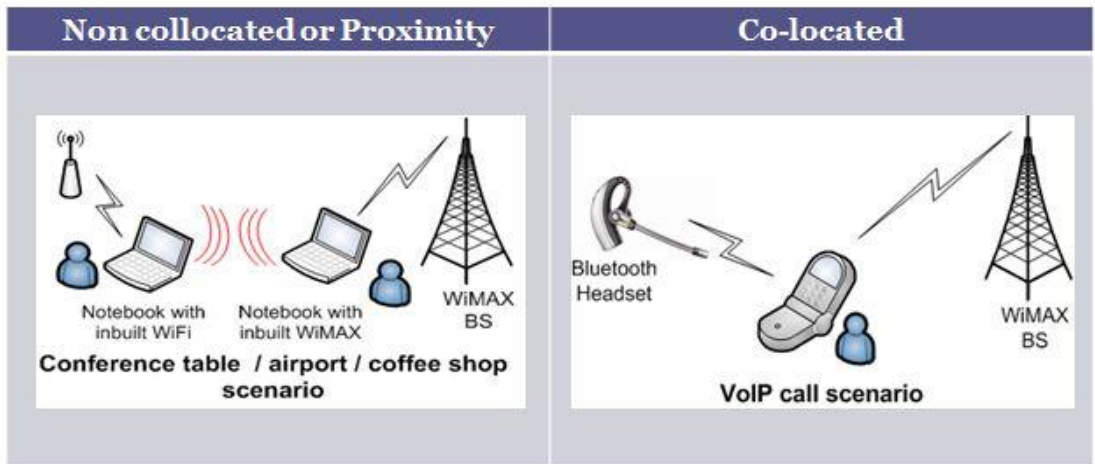

Figure 1.2: Typical WiMAX-WiFi Coexistence use case scenarios.

The bandwidth for WiFi is defined to be 22MHz in specs and for WiMAX, it can be 3.5, 5, 8.75, 10 MHz depending upon the band we are using. However the 22 MHz bandwidth for WiFi contains only the first side lobe and in reality even at a separation of 114 MHz WiFi signals can be received with signal strength of -75 dBm. Therefore in a conference room where one user chooses WiFi and other WiMAX to connect; a WiFi node can potentially interfere with nearby WiMAX node and vice-versa. Similarly in case of a multiradio platform such as latest Notebook PCs the co-located wireless interfaces may severely interfere with each other.

# Chapter 2
# Non Co-located Coexistence

## 1 Previous Works

1. Advanced filters to reduce and possibly remove interference.
2. Transmit Power Control [TPC]
3. Dynamic Frequency Selection [DFS]
4. MS led TDM [Fujitsu's Proprietary System]
5. Conservative Distributed Interframe Space approach
6. MS led demand based TDM [RTS/CTS] scheme

## 2 Motivation

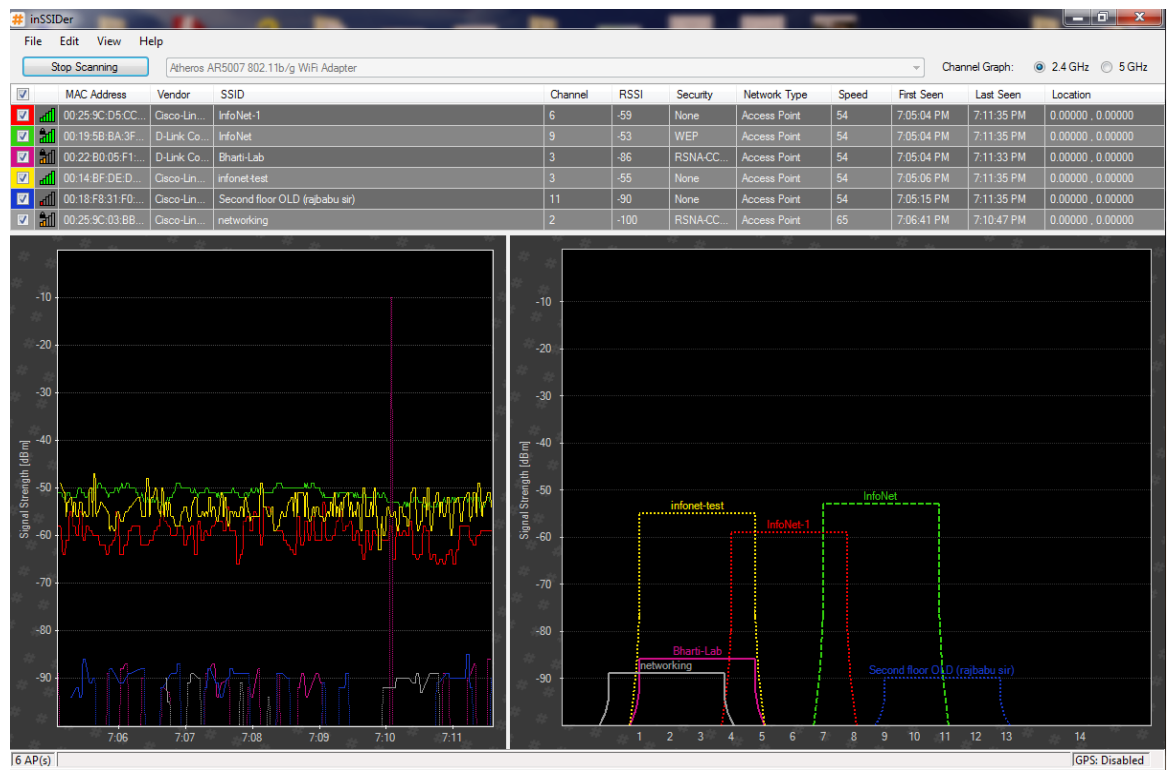

Figure 2.1: A typical view of "inSSIDer" WiFi analyzer.

We first explored that whether the interference problem between WiMAX and WiFi really exists and if it exists what can be the effects of this problem. For this purpose we used "inSSIDer wireless analyzer" in InfoNet lab to see the interference possibilities of WiFi. We found that there was really a significant possibility for both WiFi and WiMAX to interfere with each other. The results obtained by the practical experimentation with signal generator and actual WiMAX hardware at Staccato Communications gave a strong proof and precise values for this interference strength. The WiFi channel at 2.412 GHz [BG 1] generates out of band spillage of up to -61 dBm which results in an inband interference for the adjacent 2.380 GHz WiMAX channel. Similarly 2.462 GHz [BG 11] generates an in-band interference of levels up to -75 dBm for the adjacent 2.576 GHz WiMAX channel. Hence the study shows that isolation number of 57dB is required between WiMAX and WiFi antennae, which corresponds to a free space separation distance of around 7m.

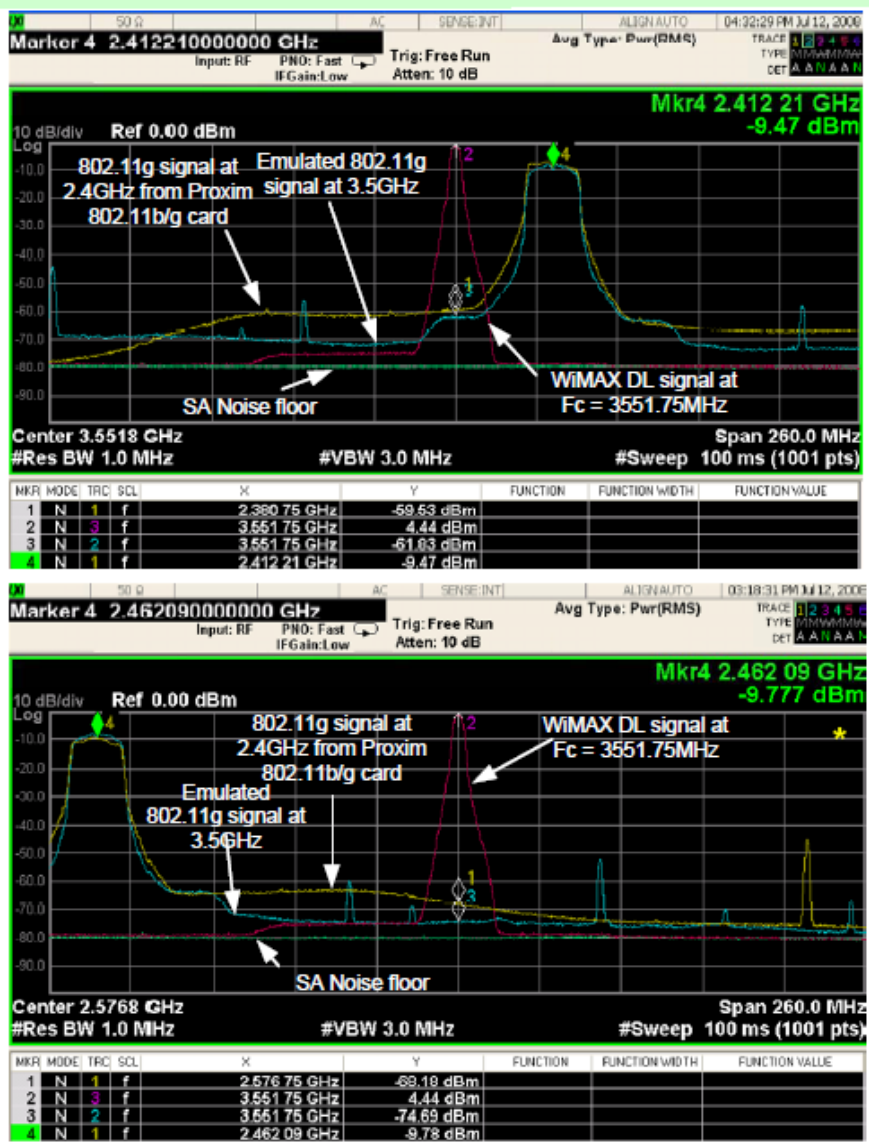

Figure 2.2: The spectrum analyzer plot. (Courtesy: Staccato Communications, San Diego, CA)

The spectrum analyzer plots also show difference in out of band emissions generated by signal generator and actual WiFi hardware (compare blue and yellow traces in plot). This suggests that the 7m distance isolation may also be a conservative estimate. The results obtained by Intel's research were also validating the severity of this interference and gave similar results with an additional fact that WiMAX node also interferes with nearby WiFi channel and isolation numbers of 60 dB and 56 dB is required respectively to prevent WiFi-WiMAX and WiMAX-WiFi interference.

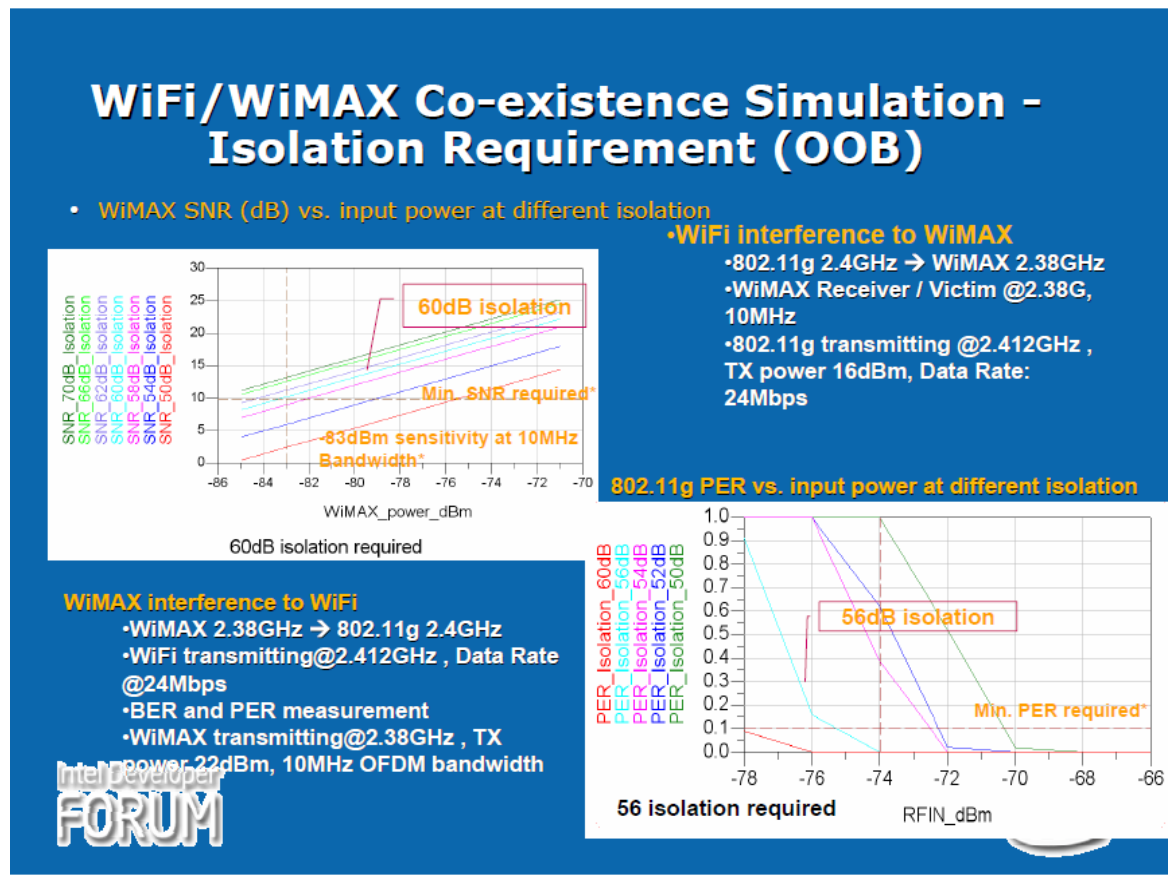

Figure 2.3: Intel Developer Forum's report.

All of the suggested approaches so far were in fact either difficult to implement or inefficient solutions in terms of overhead and fairness. We found being license-exempt, ISM band is highly popular and has billions of devices operating in it. This trend is only growing and will extend to higher bit rate applications (such as video and online gaming) which exhibit higher duty cycle. One of its application WiFi hotspot is even now being used ubiquitously. All these inferences were quite encouraging for us to work in this direction and we started thinking about an efficient method and finally came up with Adaptive Frame Reservation Scheme.

## 3 Adaptive Frame Reservation Scheme

### 3.1 Frame Reservation
We here develop a demand based time division multiplexing approach which adaptively decide best thing to do and target at optimizing the performance in terms of both throughput and fairness.

Both WiMAX (or the other 4G technology) and WiFi are most likely to be present in future multi-radio platforms such as Laptops, PDAs and Phones. A conference room (or airport) where 2 users sitting next to each other choose to connect to WiFi and WiMAX networks respectively can potentially interfere with each other. In such cases the 'unused' WiFi radio (collocated with WiMAX) can be used for coordination purpose. In WiFi MAC uses contention access where all subscribers compete for APs attention whereas WiMAX MAC uses a scheduling algorithm for which the subscriber station need to compete only for initial entry into the network and once allocated the time slot can enlarge and contract, but remains assigned to the same subscriber station.

Clearly whenever a WiMAX node is transmitting no other WiMAX node is scheduled only a WiFi node can interfere due to its random access, which may cause an erroneous WiMAX frame which would be discarded by receiver. We need to go for such a scheme where both systems can relish their own way of access to the channel without interfering with each other.

The main idea is very simple, that is, if a WiMAX node is transmitting no one else should interrupt it. The channel is free to use by any of the two technologies, however a WiMAX node before communicating with BS (either transmitting or receiving) reserves the channel to itself for a particular duration. This reservation is done by transmitting a CTS [Clear-To-Send] packet using collocated WiFi interface. The CTS is received by all the neighbors (including both WiMAX and WiFi nodes) at their WiFi interfaces and to honor 802.11 specifications all of them refrain transmitting for the duration specified in CTS. This Clear-To-Send signal transmitted by the collocated WiFi interface of a WiMAX node to itself is called Frame Reservation Signal [FRS].

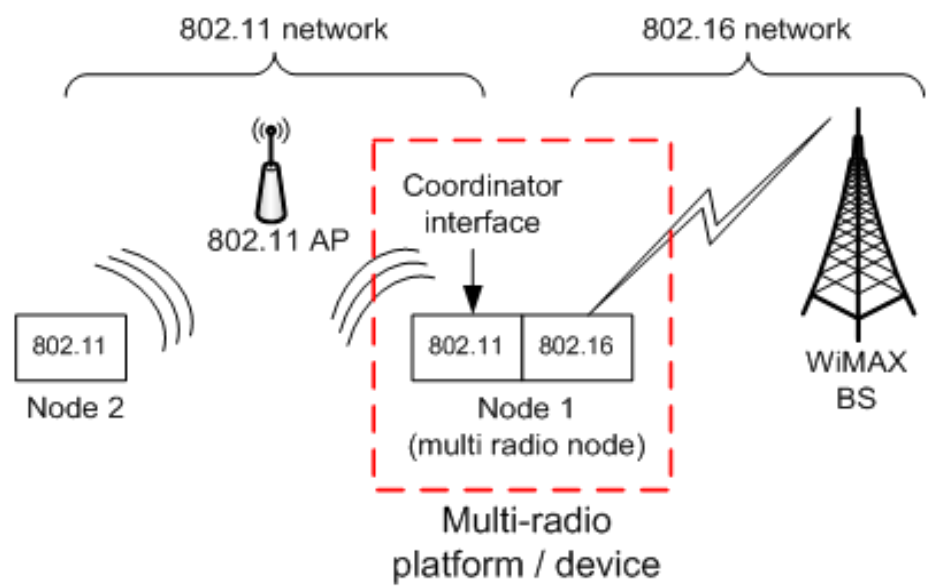

Figure 2.4: Typical WiMAX node with collocated WiFi.

In the nut shell whenever a WiMAX node needs to communicate with BS, its collocated WiFi interface generates a Frame Reservation Signal [FRS] for it, which is heard locally by all possible interferers and they all stop transmitting beforehand.

## 3.2 Performance Improvements

The BS of WiMAX system is supposed to be purely 802.16 node. There is no benefit in making it capable of using Frame Reservation Scheme because BS is supposed to serve a vast area and coexistence is a local problem that too occurring less frequently. So it is very intuitive that the WiMAX Subscriber Station [SS] itself would be enabled with collocated WiFi interface. It uses this WiFi interface for CTS transmission for Frame Reservation and to honor the CTS sent by other nodes (whether they are WiFi or WiFi enabled WiMAX).

### 3.2.1 Dynamic Medium Acquisition [DMA]

To bring fairness in this scheme we use Dynamic Medium Acquisition [DMA] where instead of being idle the WiMAX node monitors the channel whenever it is not communicating with BS and estimates how many numbers of other potential interferers are active in the neighborhood. This knowledge allows a WiMAX node to set a utilization goal (for example 33% if there is one 802.16 system and two 802.11 systems in the area) to ensure fair sharing of the medium for the deployed systems. An assessment of how much of the 33% is successfully being claimed can be used to modify the Dynamic Medium Acquisition (DMA) algorithm. The DMA algorithm sets intervals when an 802.16 system can begin monitoring and subsequently claim the medium. This interval is based on the past utilization and the utilization goal. As the utilization goal is achieved the opportunities to claim the medium are reduced.

### 3.2.2 Adaptive CTS Power [ACP]
While transmitting the CTS our purpose is to block all the interfering WiFi nodes which happen to be in the vicinity of WiMAX node. Transmitting a CTS packet with high power is likely to block almost all the WiFi nodes in the neighborhood even if they are not interfering with WiMAX. The channel clearing message can be sent at a power level equivalent to the separation distance between the interfering and the interfered nodes. Such a scheme has the advantage of restricting performance degradation to the interfering nodes only.

### 3.2.3 Dynamic Performance Evaluation [DPE]

- *If there is no Interfering WiFi node, then why unnecessarily send CTS and increase overhead?*
  Avoid CTS transmission; a large number of retransmissions occurred on WiFi or WiMAX can be an indication to switch CTS transmissions on.

- *CTS overhead and retransmission trade-off*
  If after enabling the CTS mechanism performance improves, go ahead; otherwise switch it off. Improvement here is essentially in terms of throughput not the number of collisions.

- *Frame reservation duration Threshold*
  The Frame reservation duration should not be too small or too large. Small duration increases the CTS overload and large increases average delay for WiFi. Set a Threshold Duration [$TH_{dur}$] at WiMAX node, and If demand is smaller than $TH_{dur}$ skip sending CTS packet.

- *QoS satisfaction*
  If there is some additional QoS requirement (e.g. delay, throughput) for WiMAX then those constraints can be used in Dynamic Performance Evaluation [DPE] to decide whether the scheme fulfill those requirement or not and the result can be used to give a feedback to the DMA-phase to estimate duration for which a WiMAX node reserves the medium.

## 4 Emulation

### 4.1 Test Setup

1. Remove the original ath and ath5 drivers from the LINUX kernel.
2. Install of madwifi-ng driver and lorcon library.
3. Run the LORCON code.
4. Send a CTS packet with fixed duration.
5. Put the wireless interface ath0 on the monitor mode.
6. Keep two WiFi enabled nodes nearby having ftp session going with access point as shown in figure.

### 4.2 Experiment

We conducted some practical measurements to test the impact of low-power CTS transmission. Details are furnished below:

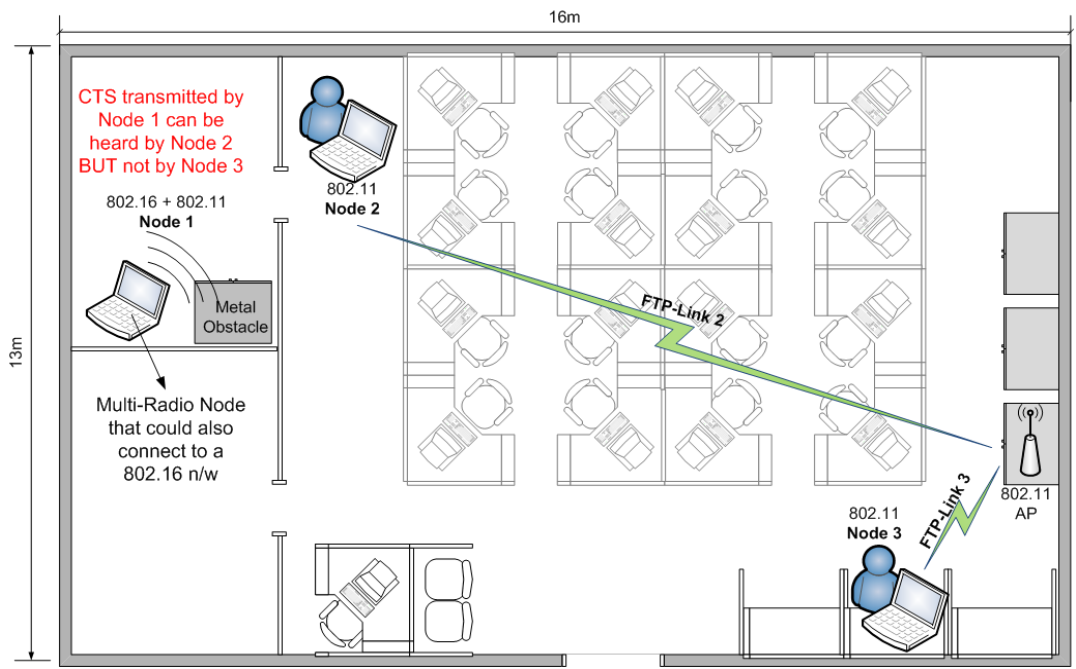

Figure 2.5: Test setup in InfoNet Lab IITB

- Two simultaneous FTP sessions (AP -> Node 2, AP -> Node 3)
- Distance between Node 1 (coordinator Interface) and Node 2 : approx. 3m
- Coordinator interface - CTS injection using madwifi+lorcon (Atheros card) with transmit power 1dBm

## 4.3 Result

FTP Link 2 was silenced by the CTS message from Node 1 however FTP Link 3 remain unaffected since Node 3 does not hear the low-power CTS message. When this CTS duration expired FTP Link 2 was again restored.

## 5 Conclusion

The AFR scheme discussed here shows a great potential to mitigate the WiMAX-WiFi Coexistence Problem. The Dynamic Medium Acquisition [DMA] and Adaptive CTS Power [ACP] put fairness and throughput improvement in the scheme. Dynamic Performance Evaluation [DPE] evaluates the performance and check the QoS fulfillment. It acts as a feedback unit and makes the scheme adaptive.

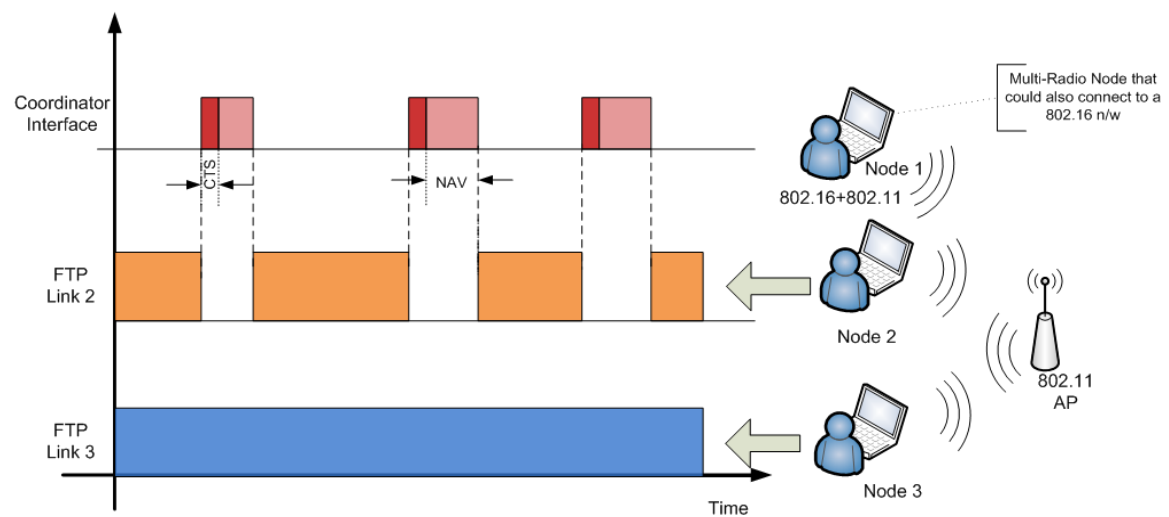

Figure 2.6: A CTS transmission pauses FTP session which is again restored after the CTS duration.

## 6 Future Work

- This work can be implemented on an actual hardware and a performance analysis in terms of both fairness and throughput can be done.
- The interference analysis towards Bluetooth and Zigbee can also be performed.

## References

[1] "Distributed Quality-of-Service Support in Cognitive Radio Networks", vorgelegt von.
[2] "UNLICENSED OPERATION OF IEEE 802.16: COEXISTENCE WITH 802.11(A) IN SHARED FREQUENCY BANDS", Lars Berlemann, Christian Hoymann, Guido Hiertz, Bernhard Walke.
[3] "WiMAX/Wi-Fi coexistence in the 3.65GHz band - standardization and simulation", Paul Piggin.
[4] "Technology for WiFi/Bluetooth and WiMAX Coexistence", Taiji Kondo, Hiroshi Fujita, Makoto Yoshida, Tamio Saito.

# Chapter 3
# Co-located Coexistence

**1 Previous Works**

1. Network coexistence
2. WiMAX scheduling
3. Modifying WiMAX base station's scheduling algorithms
4. Active Interval Allocation

**2 Motivation**

Let's consider a probable scenario in which a notebook PC is in a Local WiFi Network with wireless enabled printer, Camera, Cell Phone and other devices. At the same time it needs to access internet via WiMAX link. Thus Notebook PC needs to establish simultaneous connections to both local WiFi network and wide area WiMAX. Such a situation can pose a severe interference problem as both the radio interfaces are together on the same board. The severity is clearly more than what was explored in case of Non Co-located Scenario.

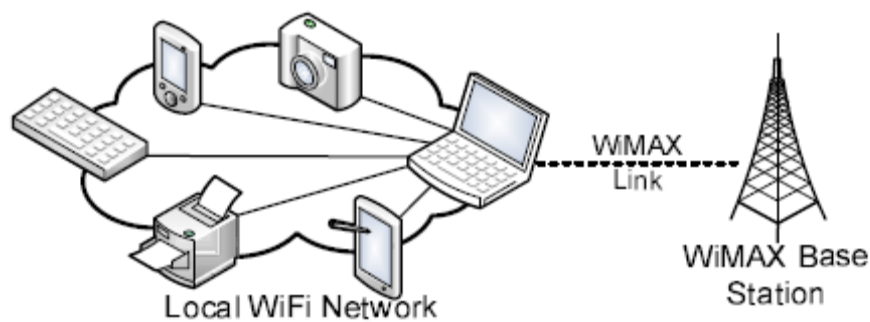

Figure3.1: A WiMAX-WiFi Coexistence Scenario with Multiradio Notebook PC as interference victim

Today's mobile devices support many wireless technologies to achieve ubiquitous connectivity. Economic and energy constraints, however, are driving the industry to implement multiple technologies into a single radio. This system-on-a-chip architecture leads to competition among networks when devices toggle across different technologies to communicate with multiple networks. An out-of-band spillage of -61dBm and -75dBm by the WiFi interface can be experienced as in-band interference by co-located WiMAX interface. Therefore a method to combat this rampant problem is needed.

# 3 Co-Located Coexistence [CLC] Controller

The co-located interference problem can be solved with the help of a simple time sharing method. As in this case multiple wireless interfaces are on a single machine we do not require Over-The-Air [OTA] Control, instead simple software based signaling can be done. Suppose a particular scenario in which two wireless interfaces (say WiMAX and Bluetooth) compete for time slots in such a case an Active Interval can be defined for time sharing purpose. Whenever collocated BT needs to communicate with another BT node in vicinity and Active Interval can be leased to it for required duration, at the end of which it has to surrender the time slots to WiMAX.

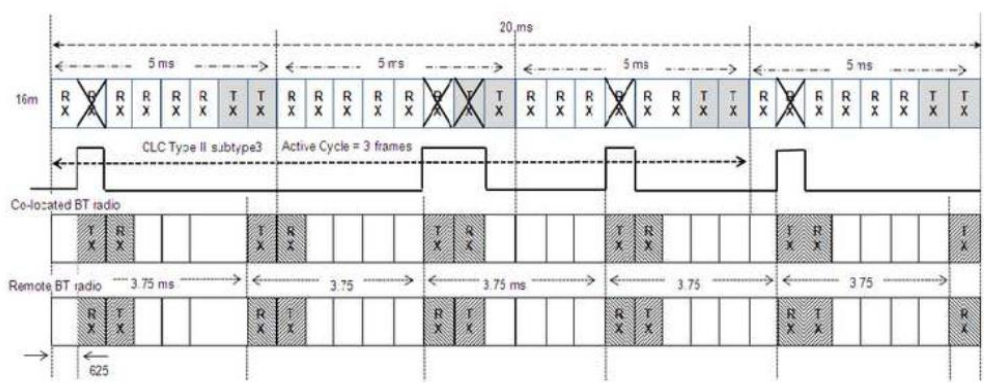

Figure 3.2: A WiMAX-BT Coexistence using Active intervals

To control and schedule various radio interfaces attached to a single Advanced Mobile Station [AMS], we intuitively need a common radio scheduler which coordinates all the interfaces simultaneously [1]. This common scheduler is Co-Located Coexistence [CLC] Controller. All radio interfaces are attached to this CLC Controller. A typical CLC Controller should have these functionalities-

- Monitor the traffic pattern of active radio modules
- Dynamic CLC activation decision
- Triggering the provisions to setup CLC procedure in the respective CLC modules of the respective radio modules.

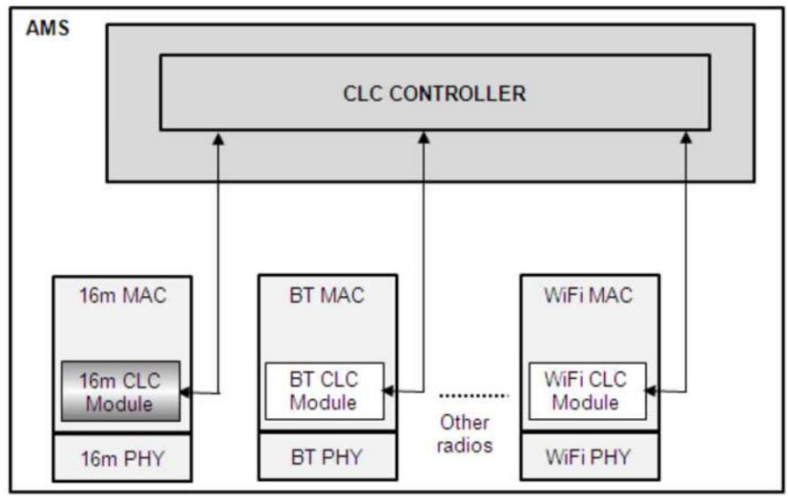

Figure 3.3: Co-Located Coexistence Controller

Co-Located interference can only occur when one of the interfaces is transmitting and another is receiving. If both the interfaces are transmitting they are not exposed to the reception of high strength spillage of another radio interface. Similarly if both interfaces are in reception mode the possible source of interference are not co-located on the same machine and thus doesn't fall in Co-located Interference category.

| WiMAX \WiFi | Tx | Rx |
|---|---|---|
| Tx | Allowed | Not Allowed |
| Rx | Not Allowed | Allowed |

Table 3.1: Transmission-Reception Case Study

Thus simultaneous transmission or reception at all the interfaces is allowed.

**4 Naïve Persistent State Controller [NPSC]**

Based upon the fundamental knowledge of Co-located Coexistence and Transmission-Reception case study, we can design a simple and naïve CLC controller. Below are the details of its fundamental blocks-

- There is a single CLC Controller 'C'.
- All the radio interfaces are attached to this CLC Controller.
- All interfaces can talk to this common CLC Controller only
- Every interface generates requests for transmission, reception and sleep mode which are sent to the CLC Controller.
- CLC Controller decides whether to accept the request or not.
- In case a request is accepted both CLC Controller and the requesting interface change their states.
- There are three states for CLC Controller [C]-
  States={0,1,2}
  0 -> S    [Sleep mode, nothing going on]
  1 -> Rx   [Reception going on]
  2 -> Tx   [Transmission going on]

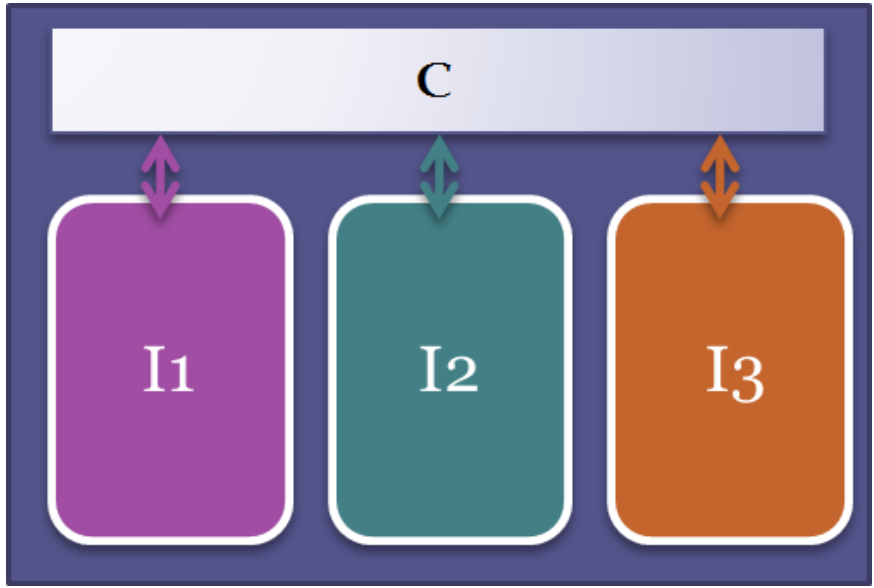

Figure 3.4: Co-Located Coexistence Controller

| C.State\I.Req | 0 [S] | 1 [Rx] | 2 [Tx] |
|---|---|---|---|
| 0 [S] | 0 | 1 | 2 |
| 1 [Rx] | 0 | 1 | X |
| 2 [Tx] | 0 | X | 2 |

Table 3.2: State Transition Table

This is a naïve scheme which completely eliminates the possibility of co-located interference. Any request from an interface that tries to persist the CLC state is always accepted. Thus the scheme is named Naïve Persistent State Controller.

Because this primitive scheme doesn't consider traffic pattern of attached radio interfaces and makes hard decision CLC activation rather than a dynamic one, It can highly effect throughput of Scheduled radio interfaces like WiMAX.

## 5 Performance Improvements

To improve upon NPSC scheme and come-up with an adaptive and smart Coexistence Controller certain refinements are needed in current scheme.

## 5.1 Traffic Pattern Information

The frames which are scheduled for the WiMAX Tx/Rx should be avoided for Co-located WiFi Rx/Tx respectively. How many subframes it uses gives information of WiMAX's share

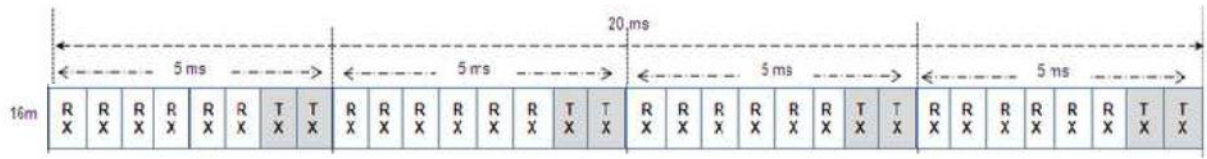

Figure 3.5: WiMAX Frame and Time Slots

## 5.2 Load Balancing
WiFi interface which monitors WiFi traffic can generate a table of other WiFi interfaces. This information can be used for fair scheduling[2].

## 5.3 Service Policies
In case of a tie priority should be given to more important or urgent task.

- WiMAX>WiFi
- Real Time Application
- QoS & Policy issue

## 5.4 Streaming Applications
It should be ensured that if a long file is to transmitted or received (e.g. ftp session, streaming), connection doesn't interrupted.